\documentclass{appolb}
\usepackage{epsfig}
\usepackage{subfigure}
\usepackage{amssymb}
\usepackage{amsfonts}
\usepackage{amsmath}

\begin{document}
\pagestyle{plain}
\title{B Physics: CP Violation beyond the SM}
\author{F.J. Botella
\address{Departament de F\'{\i}sica Te\`{o}rica and IFIC,\\ Universitat de Val\`{e}ncia-CSIC,\\ E-46100, Burjassot, Spain.}}
\maketitle

\begin{abstract}
We analyse the present experimental evidence for a complex CKM matrix, even
allowing for New Physics contributions to $\epsilon _{K}$, $a_{J/\Psi K_{S}}$%
, $\Delta M_{B_{d}}$, $\Delta M_{B_{s}}$, and the $\Delta I=1/2$ piece of $%
B\rightarrow \rho \rho $ and $B\rightarrow \rho \pi $. We emphasize the
crucial r\^{o}le played by the angle $\gamma $ in both providing irrefutable
evidence for a 3$\times $3 complex CKM matrix and placing constraints on the
size of NP contributions. It is shown that even if one allows for New
Physics a real CKM matrix is excluded at a 99.92\% C.L., and the probability
for the phase $\gamma $ to be in the interval $[-170^{\circ };-10^{\circ
}]\cup \lbrack 10^{\circ };170^{\circ }]$ is 99.7\%. Large value of the
phase $\chi $, e.g. of order $\lambda $, is only possible in models where
the unitarity of the $3\times 3$ Cabibbo-Kobayashi-Maskawa matrix is
violated through the introduction of extra $Q=2/3$ quarks. We study the
allowed range for $\chi $ and the effect of a large $\chi $ on various
low-energy observables, such as CP asymmetries in $B$ meson decays. We also
discuss the correlated effects which would be observable at high energy
colliders, like decays $t\rightarrow cZ$, etc..
\end{abstract}

\section{INTRODUCTION}

As is well-known Unitarity Triangle fits indicate the prominent role of the
Cabbibo-Kobayashi-Maskawa (CKM) mechanism\cite{CKM} in CP violation and
Flavour Physics. The huge amount and the variety of CP collected data allows
for a systematic search of Physics Beyond the Standard Model (SM)\cite{NPbotella}\cite{fits}.

A quite general and natural framework to go beyond the SM is:

\begin{enumerate}
\item Allow for New Physics in every place except in weak tree-level
dominated processes.

\item Assume $3\times 3$ unitarity.
\end{enumerate}

This framework is general enough to include practically all models of New
Physics (NP) except those that explicitly violate $3\times 3$ unitarity.
Going beyond $3\times 3$ unitarity makes the analysis almost impossible in a
model independent way\cite{silvaUni}. We will proceed in the first part with
assumptions 1 and 2.

To go on with this analysis it is very important to understand the key role
played by a measurement of the unitarity triangle phase $\gamma $. It is
very convenient to write the CKM matrix in the following way\cite{aleksan}%
\cite{GusLibr}:%
\begin{equation}
\left( 
\begin{array}{cccc}
\left\vert V_{ud}\right\vert  & \left\vert V_{us}\right\vert e^{i\chi
^{\prime }} & \left\vert V_{ub}\right\vert e^{-i\gamma } & ... \\ 
-\left\vert V_{cd}\right\vert  & \left\vert V_{cs}\right\vert  & \left\vert
V_{cb}\right\vert  & ... \\ 
\left\vert V_{td}\right\vert e^{-i\beta } & \left\vert V_{ts}\right\vert
e^{i\chi } & \left\vert V_{tb}\right\vert  & ... \\ 
... & ... & ... & ...%
\end{array}%
\right)   \label{matrix1}
\end{equation}%
Where the rephasing invariant CP violating phases\ are\cite{botella03}%
\begin{equation}
\begin{array}{ccc}
\beta =\arg \left( -V_{cd}V_{cb}^{\ast }V_{td}^{\ast }V_{tb}\right)  & , & 
\gamma =\arg \left( -V_{ud}V_{ub}^{\ast }V_{cd}^{\ast }V_{cb}\right)  \\ 
\chi =\arg \left( -V_{ts}V_{tb}^{\ast }V_{cs}^{\ast }V_{cb}\right)  & , & 
\chi ^{\prime }=\arg \left( -V_{cd}V_{cs}^{\ast }V_{ud}^{\ast }V_{us}\right) 
\end{array}
\label{phases1}
\end{equation}%
It has been shown from $3\times 3$ unitarity -$\lambda \sim 0.2$- that $\chi
^{\prime }\sim \lambda ^{4}$ and $\chi \sim \lambda ^{2}$ even outside the SM%
\cite{aguilar}. So $\chi ^{\prime }$ is too small to consider its
measurement. $\beta $ and $\chi $, accompanying $\left\vert
V_{td}\right\vert $ and $\left\vert V_{ts}\right\vert $ respectively, will
enter only in loop processes: virtual transitions $q\rightarrow t$.
Therefore processes that in the SM measure $\beta $ and $\chi $ could be
contaminated by NP. \emph{Processes that measure }$\beta $ \emph{in the SM
framework will be no longer evidence of CP violation in the CKM matrix in
the presence of NP}. The unique measurable phase that can be extracted from
tree-level processes and therefore not contaminated by NP is $\gamma $,
because it can appear in a tree-level transition $b\rightarrow u$ \cite%
{NPbotella}.

We can conclude that just $\gamma $ and the moduli of the first two rows are
the unique parameters whose extraction from experimental data is not
contaminated by NP. Or put in another way, these parameters -extracted from
weak tree-level decays- are valid in all the models included in our
assumptions 1 and 2.

In section II we will parametrize and clarify the NP physics contributions
to the contaminated flavour and CP observables. Having seven experimental
data of the CKM matrix in these general class of models, in section III we
will address the question of the dominance of the CKM mechanism as the
origin of CP violation in these models. At the same time, with the
contaminated observables, we will set bounds on the NP contributions.
Finally in section IV we will study potential large deviations of $\chi $
from its SM value in models that violates $3\times 3$ unitarity.

\section{NP CONTRIBUTIONS IN $\protect\alpha ,\protect\beta $ AND $\protect%
\gamma $ MEASUREMENTS}

As we have explained the measurements of $\beta $ will be contaminated by
NP. Also it is known that $\beta $ measurements come always from the $\beta $
contribution to $B_{d}^{0}-\overline{B}_{d}^{0}$ mixing. The simple reason
why $\beta $ contributions to $B$ decay amplitudes cannot be measured - in a
clean theoretical way- is because it always enter in the SM with a second
weak phase. Therefore it is enough to parametrize the NP contributions to
the $B_{d}^{0}-\overline{B}_{d}^{0}$ mixing. We follow reference \cite%
{branco99} to write the off diagonal mixing matrix as%
\begin{equation}
M_{12}^{\left( d\right) }=\left( M_{12}^{\left( d\right) }\right)
^{SM}r_{d}^{2}e^{-i2\phi _{d}}  \label{mixing1}
\end{equation}%
the SM corresponds to $r_{d}=1$ and $2\phi _{d}=0$. In this way we get%
\begin{equation}
\left( \frac{q}{p}\right) _{B_{d}}=\left( \frac{q}{p}\right)
_{B_{d}}^{SM}e^{2i\phi _{d}}=e^{-2i\left( \beta -\phi _{d}\right) }
\label{mixing2}
\end{equation}%
and therefore, by defining%
\begin{equation}
\overline{\beta }=\beta -\phi _{d}  \label{betabar1}
\end{equation}%
we have for the experimental observables%
\begin{eqnarray}
\Delta M_{B_{d}} &=&\left( \Delta M_{B_{d}}\right) ^{SM}r_{d}^{2}
\label{deltam1} \\
S_{J/\psi K_{S}} &=&\sin 2\left( \beta -\phi _{d}\right) =\sin 2\overline{%
\beta }  \label{sinbetabar1}
\end{eqnarray}

Clearly an independent knowledge of the full CKM matrix together with the $%
B_{d}^{0}\rightarrow J/\psi K_{S}$ asymmetry $S_{J/\psi K_{S}}$ and the mass
difference $\Delta M_{B_{d}}$ will give us the opportunity to test for the
NP parameters $r_{d}$ and $\phi _{d}$.

As we will see in the next section, the actual knowledge of the moduli of
the CKM matrix in the first two rows is not enough to know the entire CKM
matrix, so the measurement of gamma is of paramount importance in order to
complete the knowledge of the CKM matrix and to proceed with the NP
analysis. Gamma can be obtained from the phase of the rephasing invariant
quartet $V_{us}V_{cb}V_{ub}^{\ast }V_{cs}^{\ast }$ \cite{GusLibr}. A way of
measuring it from pure tree-level decays is trough the interference of the
two pure charged currents decay paths $b\rightarrow s\overline{u}c$ and $%
b\rightarrow s\overline{c}u$ \cite{gronauGamma}\cite{giri}. Babar\cite%
{babar1} and Belle\cite{belle1} have presented results using the Dalitz plot
analysis in $B^{\pm }\rightarrow DK^{\pm }$ with the subsequent decay $D^{0},%
\overline{D}^{0}\rightarrow K_{S}\pi ^{+}\pi ^{-}$. If any potential NP in $%
D^{0}-\overline{D}^{0}$ mixing is neglected\cite{amorin}, as seems
reasonable, these analysis provide a measurement of gamma free from NP
contributions.

Other relevant ways of measuring $\gamma $ are the methods to
measure\linebreak\ $\alpha =\arg \left( -V_{td}V_{tb}^{\ast }V_{ud}^{\ast
}V_{ub}\right) $ \cite{gronaualpha}. By now it is well-known that using eq.(%
\ref{phases1}) one has $\alpha =\pi -\beta -\gamma $ by definition\cite%
{silva04}\cite{barenboim97}. So a measurement of $\alpha $ is nothing else
than a measurement of $\beta +\gamma $. The main channels are $B\rightarrow
\pi \pi ,\rho \pi $ and $\rho \rho $. In this case, the presence of penguin
pollution could become NP pollution, so one has to be much more careful in
these processes. The relevant observables are%
\begin{equation}
\lambda _{f}=\left( \frac{q}{p}\right) _{B_{d}}\frac{A( \overline{B}_{d}^{0}\rightarrow f) }{A( B_{d}^{0}\rightarrow f) }
\label{lambdaf1}
\end{equation}%
To understand the effects of NP let us first neglect the SM penguin
pollution and treat the $\pi \pi $ channel as if it were a pure tree level
and therefore without NP pollution in the decay amplitudes. In these case we
have in the $\pi ^{+}\pi ^{-}$ channel%
\begin{equation}
\lambda _{+-}=e^{-i2\overline{\beta }}e^{-i2\gamma }=e^{i2\overline{\alpha }}
\label{lambdafplusminus1}
\end{equation}%
where $\overline{\alpha }\equiv \pi -\overline{\beta }-\gamma =\alpha +\phi
_{d}$. The CP asymmetry measures $\overline{\alpha }$. Once we include
penguin pollution, the Gronau and London isospin analysis can be partially
summarized with the following formula\cite{silva04}\cite{botellasilva}%
\begin{equation}
\lambda _{+0}\equiv \left( \frac{q}{p}\right) _{B_{d}}\frac{A(
B^{-}\rightarrow \pi ^{-}\pi ^{0}) }{A( B^{+}\rightarrow \pi
^{+}\pi ^{0}) }=\frac{1}{\lambda _{+-}^{\ast }}\frac{\overline{R}\pm i%
\sqrt{\overline{\rho }^{2}-\overline{R}^{2}}}{R\pm i\sqrt{\rho ^{2}-R^{2}}}
\label{lambdapluszero1}
\end{equation}%
where 
\begin{eqnarray}
\overset{{~\scriptscriptstyle\left(-\scriptscriptstyle\right)}}{R} &=&\frac{\vert \overset{\!\!\!\!\!\!{\scriptscriptstyle\left(-\scriptscriptstyle\right)}}{A_{+0}}\vert ^{2}+\frac{1}{2}\vert \overset{\!\!\!\!\!\!{\scriptscriptstyle\left(-\scriptscriptstyle\right)}}{A_{+-}}\vert ^{2}-\vert \overset{\!\!\!{\scriptscriptstyle\left(-\scriptscriptstyle\right)}}{A_{00}}\vert
^{2}}{\sqrt{2}}  \label{R1} \\
\overset{{~\scriptscriptstyle\left(-\scriptscriptstyle\right)}}{\rho } &=&\vert \overset{\!\!\!\!\!\!{\scriptscriptstyle\left(-\scriptscriptstyle\right)}}{A_{+0}}\vert\vert\overset{\!\!\!\!\!\!{\scriptscriptstyle\left(-\scriptscriptstyle\right)}}{A_{+-}}\vert
\label{Rho1}
\end{eqnarray}%
that tell us that $\lambda _{+0}$ can be extracted from the experimental
data including branching ratios and $\lambda _{+-}$. But $A_{+0}$ in the SM
and therefore in our NP scenario is a pure tree-level amplitude with weak
phase gamma - is proportional to the $\Delta I=3/2$ piece-, so we will have
that the observable $\lambda _{+0}$ will be%
\begin{equation}
\lambda _{+0}=e^{-i2\overline{\beta }}e^{-i2\gamma }=e^{-2i\overline{\alpha }%
}  \label{lambdapluszero2}
\end{equation}%
We conclude that the usual way of measuring $\alpha $ in $\pi \pi $ decays
provides us with a measurement of $\overline{\alpha }=\pi -\overline{\beta }%
-\gamma $ even in the presence of NP in the $\Delta I=1/2$ piece\cite{baek1}%
. Obviously the knowledge of $\overline{\beta }$ from $J/\psi K_{S}$
converts these $\alpha $ methods in another way of extracting $\gamma $ from
tree-level pieces. Similar results are obtained for $\rho \pi $ and $\rho
\rho $.

\section{IS THE CKM MATRIX COMPLEX IN\ THE PRESENCE OF NP?}

Within this class of models, in order to investigate whether the present
experimental data already implies that CKM is complex, one has to check
whether any of the unitarity triangles is constrained by data to be
non-\textquotedblleft flat\textquotedblright , i.e. to have a non-vanishing
area. If any one of the triangles does not collapse to a line, no other
triangle will collapse, due to the remarkable property that all the
unitarity triangles have the same area. This property simply follows from
unitarity of the 3$\times $3 CKM matrix. The universal area of the unitarity
triangles gives a measurement of the strength of CP violation mediated by a $%
W$-interaction and can be obtained from four independent moduli of $V_{CKM}$%
. The fact that one can infer about CP violation from the knowledge of
CP-conserving quantities should not come as a surprise\cite{botellaChau} .
It just reflects the fact that the strength of CP violation is given by the
imaginary part of a rephasing invariant quartet\cite{jarlskog} , $J=\pm 
\text{Im}\left( V_{i\alpha }V_{j\beta }V_{i\beta }^{\ast }V_{j\alpha }^{\ast
}\right) $, with $\left( i\neq j,\alpha \neq \beta \right) $, which in turn
can be expressed in terms of moduli, thanks to 3$\times $3 unitarity.
Restricting ourselves to the first two rows of $V_{CKM}$, to avoid any
contamination from NP, a possible choice of independent moduli would be $%
\left\vert V_{us}\right\vert $, $\left\vert V_{cb}\right\vert $, $\left\vert
V_{ub}\right\vert $ and $\left\vert V_{cd}\right\vert $. One can then use
unitarity of the first two rows to evaluate $J$, which is given, in terms of
the input moduli\cite{NPbotella}, by 
\begin{eqnarray}
4J^{2} &=&4\left( 1-\left\vert V_{ub}\right\vert ^{2}-\left\vert
V_{us}\right\vert ^{2}\right) \left\vert V_{ub}\right\vert ^{2}\left\vert
V_{cd}\right\vert ^{2}\left\vert V_{cb}\right\vert ^{2}-  \notag \\
&&-\left( \left\vert V_{us}\right\vert ^{2}-\left\vert V_{cd}\right\vert
^{2}+\left\vert V_{cd}\right\vert ^{2}\left\vert V_{ub}\right\vert
^{2}-\left\vert V_{cb}\right\vert ^{2}\left\vert V_{ub}\right\vert
^{2}-\left\vert V_{cb}\right\vert ^{2}\left\vert V_{us}\right\vert
^{2}\right) ^{2}  \label{J1}
\end{eqnarray}%
Note that Eq. (\ref{J1}) is exact, but the actual extraction of $J$ from the
chosen input moduli, although possible in principle, it is not feasible in
\textquotedblleft practice\textquotedblright .

\begin{figure}[hbt]
\begin{center}
\epsfig{file=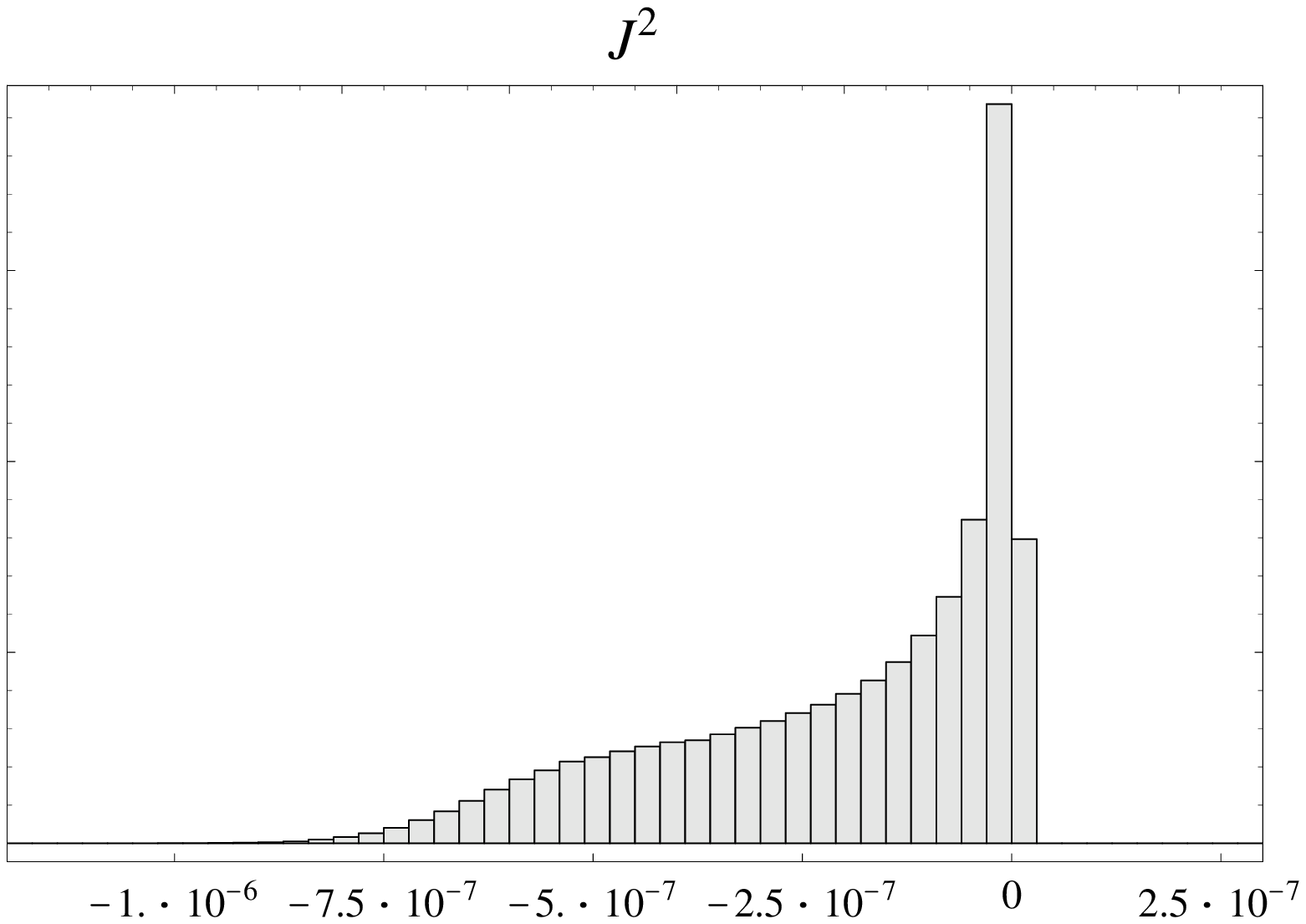,width=0.65\textwidth}
\caption{$J^{2}$ distribution from $\left\vert V_{us}\right\vert =0.2200\pm 0.0026$, $ \left\vert V_{ub}\right\vert =\left( 3.67\pm 0.47\right) \times 10^{-3}$, $\left\vert V_{cb}\right\vert =\left( 4.13\pm 0.15\right) \times 10^{-2}$ and 
$\left\vert V_{cd}\right\vert =0.224\pm 0.012$.}\label{Jpdf}
\end{center}
\end{figure}

To illustrate this point, let us consider the present experimental values of 
$\left\vert V_{us}\right\vert ,\left\vert V_{cb}\right\vert ,\left\vert
V_{ub}\right\vert $ and $\left\vert V_{cd}\right\vert $, assuming Gaussian
probability density distributions around the central values. We plot in Fig.(\ref{Jpdf}) the probability density distribution of $J^{2}$, generated
using a toy Monte Carlo calculation. Only $31.1\%$ of the generated points
satisfy the trivial normalization constraints and, among those, only $7.9\%$
satisfy the condition that the unitarity triangles close $\left(
J^{2}>0\right) $.

\begin{figure}[tb]
\begin{center}
\subfigure[Apex of the unitarity triangle $bd$.\label{gamma0}]{\epsfig{file=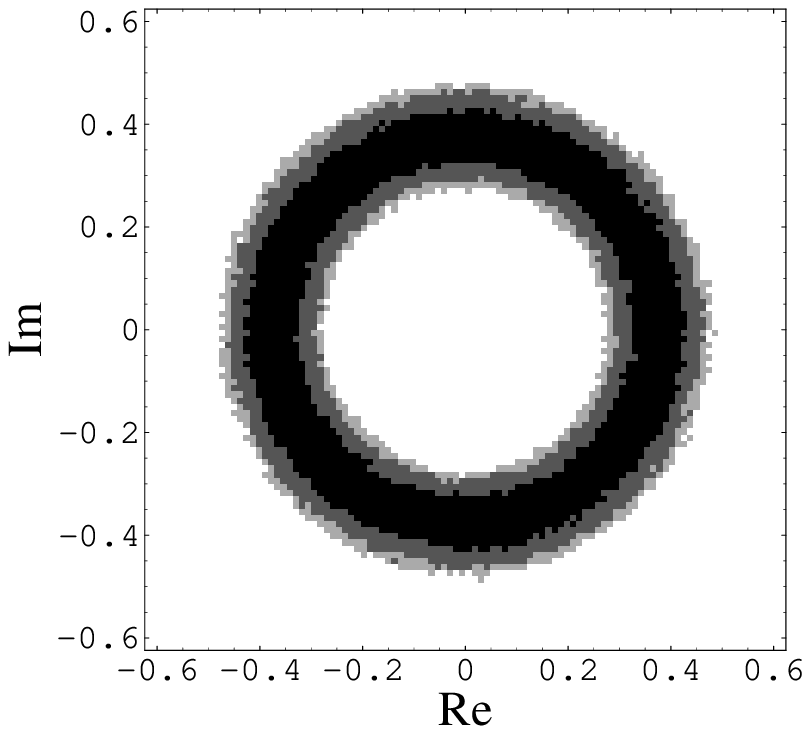,width=0.45\textwidth}}~\subfigure[$\left( r_{d}^{2},2\protect\phi _{d}\right) $ joint distribution.\label{rfi0}]{\epsfig{file=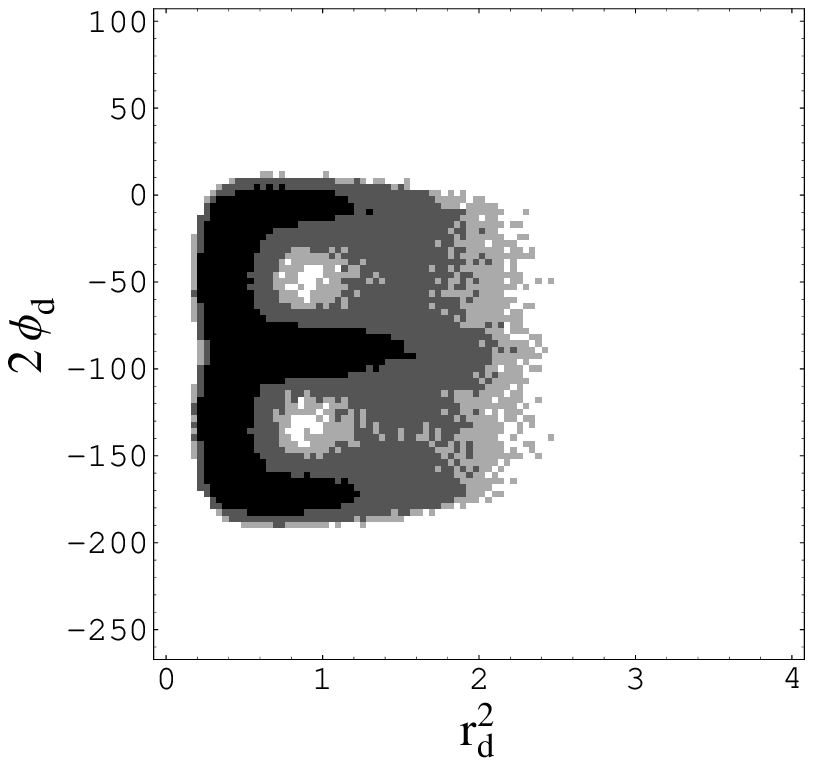,width=0.45\textwidth}}
\caption{Probability distributions, no restriction on $\gamma$.}
\end{center}
\end{figure}

As we have mentioned, including the experimental data $S_{J/\psi K_{S}}$ and $%
\Delta M_{B_{d}}$ still does not give evidence of a complex $V_{CKM}$ . We
cannot conclude if $V_{CKM}$ is the dominant contribution to CP violation,
although we can set bounds on the NP parameters $r_{d}$ and $\phi _{d}$. In
Fig.(\ref{gamma0}) we plot 68\% (black), 90\% (dark grey) and 95\% (grey)
probability regions of the probability density function (PDF) of the apex $%
-V_{ud}V_{ub}^{\ast }/V_{cd}V_{cb}^{\ast }$ of the $db$ unitarity triangle.
In Fig.(\ref{rfi0}) we represent joint PDF regions in the plane $(r_{d}^{2}$%
, $2\phi _{d})$. Because $\gamma $ gives the apex of the triangle, it is
clear from Fig.(\ref{gamma0}) that there is essentially no restriction on $%
\gamma $. On the contrary, because the moduli of the first two rows put an
upper bound on $|\beta |$ and upper and lower bounds on $R_{t}=\left\vert
V_{td}V_{tb}^{\ast }\right\vert /\left\vert V_{cd}V_{cb}^{\ast }\right\vert $%
, we can see in Fig.(\ref{rfi0}) significant constraints on $2\phi _{d}$ and 
$r_{d}^{2}$.

Including the measurement of gamma in Fig.(\ref{gamma0}) will fix the
unitarity triangle and the $V_{CKM}$ matrix. The enormous effort developed
at the B-factories Belle and BaBar has resulted in the first measurements of 
$\gamma $ in tree-level decays $B^{\pm }\rightarrow DK^{\pm }$, $B^{\pm
}\rightarrow D^{\ast }K^{\pm }\rightarrow \left( D\pi ^{0}\right) K^{\pm }$,
where the two paths to $D^{0}$ or $\bar{D}^{0}$ interfere in the common
decay channel $\bar{D}^{0},{D}^{0}\rightarrow K_{S}\pi ^{+}\pi ^{-}$. From a
Dalitz-plot analysis , Belle \cite{belle1} has presented $\gamma =68^{\circ
}\pm 
\begin{smallmatrix}
14^{\circ } \\ 
15^{\circ }%
\end{smallmatrix}%
\pm 13^{\circ }\pm 11^{\circ }$ and BaBar \cite{babar1} $\gamma =70^{\circ
}\pm 26^{\circ }\pm 10^{\circ }\pm 10^{\circ }$, together with the solutions
obtained by changing $\gamma \rightarrow \gamma \pm \pi $.

We average conservatively both measurements to the value $\gamma =69^{\circ
}\pm 21^{\circ }$ ($-111^{\circ }\pm 21^{\circ }$), which we take as a
quantitative measurement of a complex CKM matrix \emph{independent of the
presence of NP at the one-loop weak level}.

BaBar has also presented a time-dependent analysis of the $\rho ^{+}\rho ^{-}
$ channel \cite{babarRhRho}, that once supplemented with the $\rho ^{+}\rho
^{0}$ and $\rho ^{0}\rho ^{0}$ branching ratios \cite{babarBrRho,belleBrRho} and the measurement of the final polarization, can be translated 
\cite{alphaNumber} into the measured value $\overline{\alpha }=96^{\circ
}\pm 10^{\circ }\pm 5^{\circ }\pm 11^{\circ }$ where the last error comes
from the usual $SU(2)$ isospin bounds\footnote{In our notation $\overline{\alpha }_{eff}$ is the usual $\alpha _{eff}$ \
but where we have introduced $\overline{\beta }$ instead of $\beta $ as the
phase in the $B_{d}^{0}$--$\bar{B}_{d}^{0}$ mixing.} $\left\vert \overline{%
\alpha }_{eff}-\overline{\alpha }\right\vert \leq 11^{\circ }$. Because the
measurement is sensitive to $\sin (2\overline{\alpha }_{eff})$, $\overline{%
\alpha }=\overline{\alpha }_{eff}\pm 11^{\circ }$ presents a fourfold
ambiguity $\left( \overline{\alpha },\overline{\alpha }+\pi ,\frac{\pi }{2}-%
\overline{\alpha },-\overline{\alpha }-\frac{\pi }{2}\right) $. In the $\rho
\pi $ channel the pentagon isospin analysis -- from quasi-two-body decays --
needs more statistics and/or additional assumptions. A time-dependent
Dalitz-plot analysis in the channel $B\rightarrow \pi ^{+}\pi ^{-}\pi ^{0}$
has been presented by BaBar \cite{alphaRhoPi}, with the result $\overline{%
\alpha }=113^{\circ }\pm 
\begin{smallmatrix}
27^{\circ } \\ 
17^{\circ }%
\end{smallmatrix}%
\pm 6^{\circ }$.Since this analysis is sensitive to both $\sin \left( 2%
\overline{\alpha }_{eff}\right) $ and $\cos \left( 2\overline{\alpha }%
_{eff}\right) $, the resulting ambiguity is just a twofold one $\left( 
\overline{\alpha },\overline{\alpha }+\pi \right) $. It is remarkable that
these two solutions are in good agreement with two of the solutions coming
from the $\rho \rho $ channel. This important property will be used to
eliminate two of the four solutions coming from the $\rho \rho $ channel.

The situation in the $\pi \pi $ channel does not yet allow a full isospin
analysis and the isospin bounds are quite poor. Furthermore BaBar and Belle
measurements are still in some conflict, so that we will not use these
results.

As before, we average the data from $\rho \rho $ and $\rho \pi $ but only
keep the two solutions consistent with the $\rho \pi $ channel data. Our
averaged values are $\overline{\alpha }=100^{\circ }\pm 16^{\circ }$, ($%
-80^{\circ }\pm 16^{\circ }$).

To analyze the implications for the dominance of the CKM mechanism for CP
violation and the presence of NP, we add both measurements $\gamma $ and $%
\overline{\alpha }$ to the previous analysis presented in Figs.(\ref{gamma0}%
) and (\ref{rfi0}).

\begin{figure}
\begin{center}
\subfigure[Apex of the unitarity triangle $bd$.\label{gamma1}]{\epsfig{file=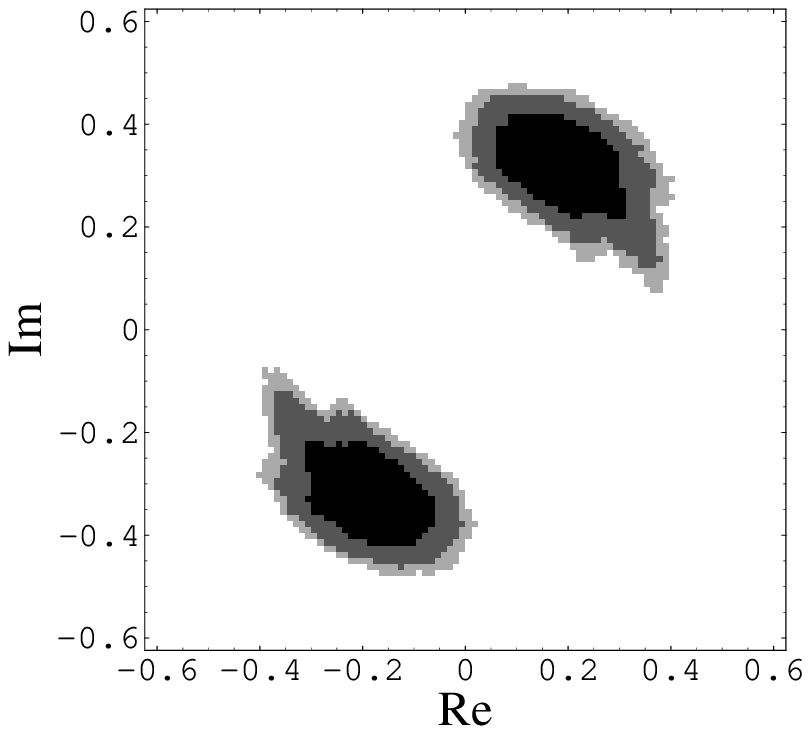,width=0.45\textwidth}}~\subfigure[$\left( r_{d}^{2},2\protect\phi _{d}\right) $ joint distribution.\label{rfi1}]{\epsfig{file=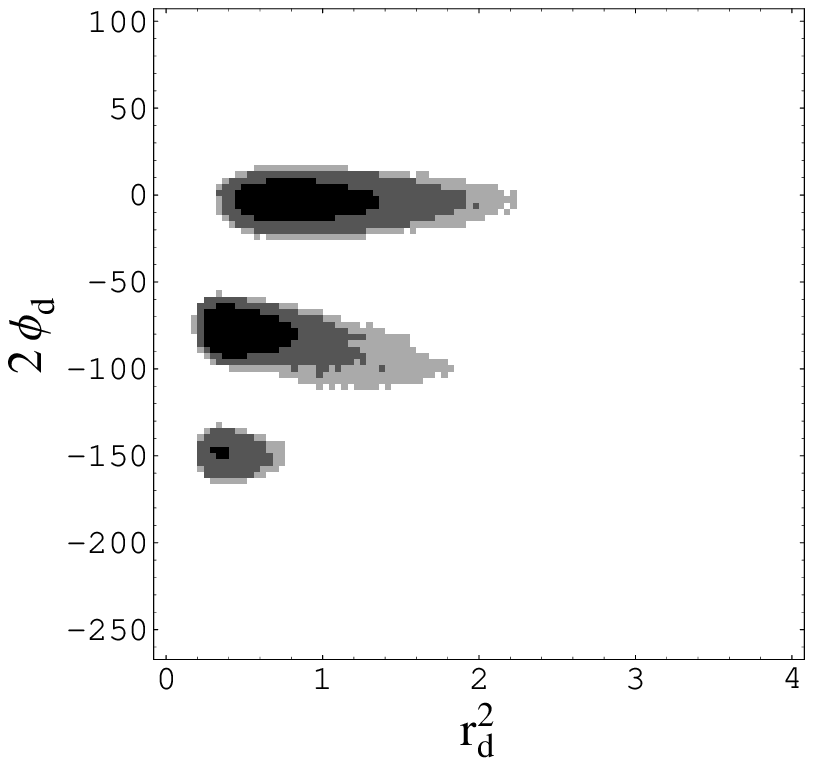,width=0.45\textwidth}}
\caption{Probability distributions including $\protect\gamma $ and $\overline{\protect\alpha }$ experimental data.}
\end{center}
\end{figure}

In Fig.(\ref{gamma1}) we represent the analogue of Fig.(\ref{gamma0}). We
conclude \cite{NPbotella} that a real CKM matrix is excluded at a 99.92\%
C.L. and the probability of $\gamma \in \lbrack 10^{\circ };170^{\circ
}]\cup \lbrack -170^{\circ };-10^{\circ }]$ is 99.7\%.

In Fig(\ref{rfi1}) we can
see three solutions in the $\left( r_{d}^{2},2\phi _{d}\right) $ plane with $%
2\phi _{d}\sim 0%
{{}^o}%
,-75%
{{}^o}%
$ and $-150%
{{}^o}%
$. In these plots one has in general four solutions corresponding to the two
values of $\gamma $ ($\overline{\alpha }$) and to the two signs of $\cos
\left( 2\overline{\beta }\right) $ The last solution corresponds to $\cos
\left( 2\overline{\beta }\right) <0$. It is the inclusion of both $\gamma $
and $\overline{\alpha }$ constraints that almost eliminates the $\cos \left(
2\overline{\beta }\right) <0$ solutions. The first solution is obviously the
SM one, and the semileptonic asymmetry $A_{SL}$ - not included here - is
starting to play an important role\cite{UTfitter1} in the exclusion of the
solution $2\phi _{d}\sim -75%
{{}^o}%
$.



It is important to stress that having an irrefutable piece of evidence for a
complex CKM matrix, in a framework where the presence of NP is allowed, has
profound implications for models of CP violation. In the particular case of
models with spontaneous CP violation, a complex CKM matrix favours the class
of models where, although Yukawa couplings are real, the vacuum phase
responsible for spontaneous CP violation also generates CP violation in
charged-current weak interactions. Conversely, the evidence for a complex
CKM matrix, even allowing for the presence of NP, excludes the class of
models with spontaneous CP violation and a real CKM matrix at a 99.92\% C.L..

\section{THE SIZE OF $\protect\chi =\arg \left( -V_{ts}V_{tb}^{\ast
}V_{cs}^{\ast }V_{cb}\right) $ AND DEVIATION FROM $3\times 3$ UNITARITY}

Within the SM and any extension where $V_{3\times 3}$ is unitary, like
supersymmetric or multi Higgs doublet models, we have the relation\cite%
{botella03} 
\begin{equation}
\sin \chi =\frac{|V_{ub}||V_{us}|}{|V_{cb}||V_{cs}|}\sin (\gamma +\chi
^{\prime }-\chi )  \label{sinChi1}
\end{equation}%
which shows that $|\chi |\lesssim \lambda ^{2}$ in any model where $3\times
3 $ CKM unitarity holds. In particular, within the SM one obtains at 90\% CL 
\begin{equation}
0.015\leq \chi \leq 0.022\quad \mathrm{(SM)}  \label{ChiSM1}
\end{equation}%
The only models in which $\chi $ can be significantly larger than $\lambda
^{2}$ are those in which $V_{3\times 3}$ is not unitary, what can only be
achieved by enlarging the quark sector. The most simple way of doing this is
with the introduction of new quark singlets \cite{singlets1}\footnote{%
The addition of a sequential fourth generation is another possibility, but
it is disfavoured by two facts: (\emph{i\/}) the experimental value of the
oblique correction parameters only leave a small range for the masses of the
new quarks; (\emph{ii\/}) anomaly cancellation requires the introduction of
a new lepton doublet, in which the new neutrino should be very heavy, in
contrast with the small masses of the presently known neutrinos.} .Quark
singlets often arise in grand unified theories and models with extra
dimensions at the electroweak scale \cite{singlets2}. They have both their
left- and right-handed components transforming as singlets under $SU(2)_{L}$%
, thus their addition to the SM particle content does not spoil the
cancellation of triangle anomalies. In these models, the charged and neutral
current terms of the Lagrangian in the mass eigenstate basis are%
\begin{eqnarray}
{\mathcal{L}}_{W} &=&-\frac{g}{\sqrt{2}}\,\bar{u}_{L}\gamma ^{\mu
}Vd_{L}\,W_{\mu }^{+}+\mathrm{h.c.}  \label{LagrangCC1} \\
{\mathcal{L}}_{Z} &=&-\frac{g}{2c_{W}}\left( \bar{u}_{L}\gamma ^{\mu }Xu_{L}-%
\bar{d}_{L}\gamma ^{\mu }Ud_{L}-2s_{W}^{2}J_{em}^{\mu }\right) Z_{\mu }
\label{lagrancNC1}
\end{eqnarray}%
where $u=(u,c,t,T,\dots )$ and $d=(d,s,b,B,\dots )$, $V$ denotes the
extended CKM matrix and $X=VV^{\dagger }$, $U=V^{\dagger }V$ are hermitian
matrices. $X$ and $U$ are not necessarily diagonal and thus flavour-changing
neutral couplings (FCNC) exist at the tree level, although they are
naturally suppressed by the ratio of the standard quark over the heavy
singlet masses. Moreover, the diagonal $Zqq$ couplings, which are given by
the diagonal entries of $X$ and $U$ plus a charge-dependent term, are also
modified. Within the SM $X_{uu}=X_{cc}=X_{tt}=1$, $X_{qq^{\prime }}=0$ for $%
q\neq q^{\prime }$, $U_{dd}=U_{ss}=U_{bb}=1$ and $U_{qq^{\prime }}=0$ for $%
q\neq q^{\prime }$. The addition of up-type $Q=2/3$ singlets modifies the
first two of these equalities, while the addition of down-type $Q=-1/3$ ones
modifies the last two.

In models with a down quark singlet, from orthogonality of the second and
third columns of $V$, one obtains the generalization of Eq.(\ref{sinChi1})%
\begin{equation}
\sin \chi =\frac{|V_{ub}||V_{us}|}{|V_{cb}||V_{cs}|}\sin (\gamma +\chi
^{\prime }-\chi )-\frac{\text{Im}\left( U_{bs}e^{-i\chi }\right) }{%
|V_{cb}||V_{cs}|}  \label{sinChi2}
\end{equation}%
From the present bound on $b\rightarrow s\ell ^{+}\ell ^{-}$, one obtains%
\cite{fitsVL}\cite{Aguilar fit} that at most $|U_{bs}|\simeq 10^{-3}\sim
\lambda ^{4}$, thus implying that in this class of models $\chi $ cannot be
significantly larger than $\lambda ^{2}$.

In model with an up quark singlet, from orthogonality of the second and
third rows of $V$, one gets\cite{aguilar}%
\begin{equation}
\sin \chi =\frac{\mathrm{Im}\;X_{ct}}{|V_{cs}||V_{ts}|}+O(\lambda ^{2})
\label{sinChi3}
\end{equation}%
In contrast with models containing down-type singlets, where the size of all
FCNC is very restricted by experiment, present limits on $X_{ct}$ are rather
weak. The most stringent one, $|X_{ct}|\leq 0.41$ with a 95\% CL, is derived
from the non-observation of single top production at LEP, in the process $%
e^{+}e^{-}\rightarrow t\bar{c}$ and its charge conjugate. This bound does
not presently provide an additional restriction on the size of $\chi $. In
models with extra up singlets $|X_{ct}|$ can be of order $\lambda ^{3}$,
yielding $\chi \sim \lambda $. From Eq.(\ref{sinChi3}), one derives some
important phenomenological consequences. First, we observe that a sizeable $%
\chi $ is associated to a FCNC $X_{ct}\sim 10^{-2}$, which \emph{leads to
FCNC decays }$t\rightarrow cZ$\emph{\ at rates observable at LHC}. In
addition, the modulus of $X_{ct}$ obeys the equality \cite{prlpaco}%
\begin{equation}
|X_{ct}|^{2}=(1-X_{cc})(1-X_{tt})  \label{Xct1}
\end{equation}%
This relation shows that conditions for achieving $X_{ct}\sim 10^{-2}$ are
to have a small deviation $O(\lambda ^{4})$ of $X_{cc}$ from unity (which is
allowed by the measurement of $R_{c}$ and $A_{FB}^{0,c}$ \cite{Aguilar fit})
and a deviation of $X_{tt}\sim \left\vert V_{tb}\right\vert ^{2}$ from unity
of order $\lambda ^{2}$. This deviation is only possible if the mass of new
top quark $T$ is bellow $1$ TeV, again testable at LHC. Finally the $D^{0}-%
\overline{D}^{0}$ mass difference sets bounds on $X_{uc}$ that for large $%
\chi $ - $(1-X_{cc})\sim O(\lambda ^{4})$ - translate into bounds on $%
(1-X_{uu})$ in such a way that deviations of unitarity in the first row are
not observable. By the same token, large values of $\chi $ will be
correlated with an important contribution to $D^{0}-\overline{D}^{0}$ mixing 
\cite{aguilar}.

\vspace{0.75cm}
\begin{figure}[htb]
\begin{center}
\epsfig{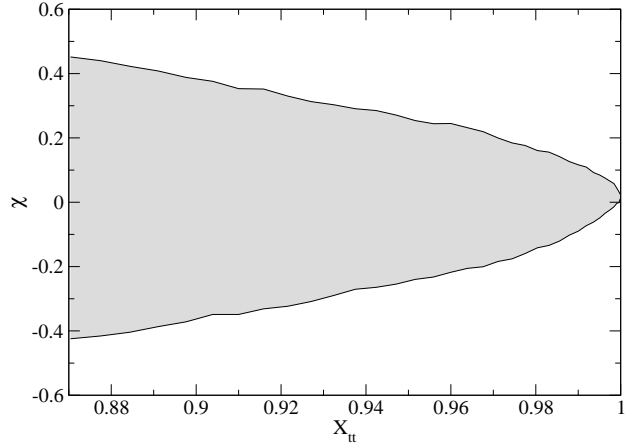}
\vspace{0.4cm}
\caption{Allowed interval of $\protect\chi $ (shaded area) as function of $X_{tt}$}\label{RangeChi}
\end{center}
\end{figure}

The shaded area in Fig.(%
\ref{RangeChi}) represents the allowed interval of $\chi $ for a given $%
X_{tt}$. Note that lower values of $X_{tt}$ are allowed for lighter $T$
quarks\cite{Aguilar fit}. The Fig.(\ref{RangeChi}) has been obtained
incorporating all the relevant constraints: the correction to the oblique
parameter $\Delta T$, $R_{c}$ and $A_{FB}^{\left( 0,c\right) }$ in the charm
sector, $D^{0}-\overline{D}^{0}$ mixing and several bounds from rare $K$ and 
$B$ decays.

Because $\chi $ is the phase of $V_{ts}$, important effects will appear in $%
b\rightarrow s$ transition in the piece $V_{ts}V_{tb}^{\ast }$, nevertheless
the presence of the new $T$ quark will introduce a contribution proportional
to $V_{Ts}V_{Tb}^{\ast }$ and dependent on the mass $m_{T}$ of this quark.
For the CP asymmetry of $B_{d}^{0}\rightarrow \phi K_{S}$ we get\cite%
{aguilar}%
\begin{equation}
S_{\phi K_{S}}=\sin (2\bar{\beta}+2\bar{\chi})  \label{sinePhyK1}
\end{equation}%
where%
\begin{equation}
\bar{\chi}=\chi -\frac{1}{2}\arg \left( \frac{1+f(m_{T},m_{t})V_{Tb}V_{Ts}^{%
\ast }/V_{tb}V_{ts}^{\ast }}{1+f(m_{T},m_{t})V_{Tb}^{\ast
}V_{Ts}/V_{tb}^{\ast }V_{ts}}\right)   \label{Chibar1}
\end{equation}%
gets contributions from $\chi $ and the new $T$ quark. $f(m_{T},m_{t})$ is
in general a complex function fixed by Wilson coefficients and matrix
elements\cite{aguilar}. By scanning all the allowed range of parameters we
get $S_{\phi K_{S}}\in \lbrack 0.57,0.93]$.

Less dependent on hadronic matrix elements is the contribution to $B_{s}^{0}-%
\overline{B}_{s}^{0}$ mixing:%
\begin{equation}
M_{12}^{B_{s}}=K\sum_{i,j=t,T^{\prime }}(V_{is}^{\ast }V_{ib})(V_{js}^{\ast
}V_{jb})S(m_{i},m_{j})=KS(m_{t},m_{t})|V_{ts}|^{2}|V_{tb}|^{2}r_{s}^{2}e^{-2i\chi _{eff}}
\label{M121}
\end{equation}%
\begin{eqnarray}
r_{s}^{2}e^{-2i\chi _{eff}} &=&e^{-2i\chi }\left\{ \left[ 1+\frac{%
S(m_{t},m_{T})V_{Ts}^{\ast }V_{Tb}}{S(m_{t},m_{t})V_{ts}^{\ast }V_{tb})}%
\right] ^{2}\right.  \notag \\
&&\left. +\left[ \frac{S(m_{T},m_{T})}{S(m_{t},m_{t})}-\left( \frac{%
S(m_{t},m_{T})}{S(m_{t},m_{t})}\right) ^{2}\right] \left( \frac{V_{Ts}^{\ast
}V_{Tb}}{V_{ts}^{\ast }V_{tb}}\right) ^{2}\right\}  \label{ChiEffec1}
\end{eqnarray}%
with $K$ a constant factor and $S$ the usual Inami-Lim box function\cite%
{InamLim}. In any channel without a weak phase in the decay amplitude, for
example in the $B_{s}^{0}\rightarrow D_{s}^{+}D_{s}^{-}$ and $\psi \,\phi $
channels, the time dependent CP asymmetry is%
\begin{equation}
S_{D_{s}^{+}D_{s}^{-}}=\sin 2\chi _{eff}  \label{sineDD}
\end{equation}%
again $\chi _{eff}$ is equal to $\chi $ plus a $T$ dependent contribution.
The range of both contributions goes in opposite direction as $m_{T}$
changes.

\vspace{0.55cm}
\begin{figure}[htb]
\begin{center}
\epsfig{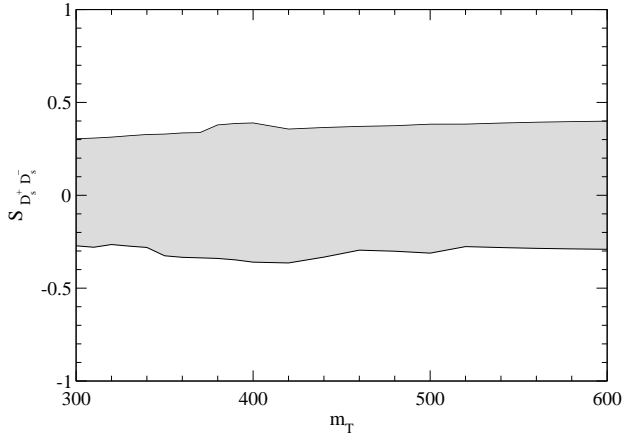}
\caption{Range of variation of the asymmetry $S_{D_{s}^{+}D_{s}^{-}}$ in terms of $m_{T}$}\label{Sdd}
\end{center}
\end{figure}

In Fig.(\ref{Sdd}), we
present the range of variation of $S_{D_{s}^{+}D_{s}^{-}}$ in terms of $%
m_{T} $. A potential spectacular departure from the SM could be seen at LCH.

\section{CONCLUSIONS}

The first measurements of $\gamma $ clearly points towards the CKM mechanism
as the dominant source of CP violation in the quark sector, even in the
presence of NP in all loops. A real CKM matrix is excluded at the 99.92\%
C.L. Including $\overline{\alpha }$ data, there are still alive three
solutions: the most robust is the SM, one with $\cos \left( 2\overline{\beta 
}\right) <0$ is almost excluded and it remains another NP solution where the
semileptonic asymmetry can do a relevant job. Future improvements of the
data will be crucial to left just very small deviations of the SM.

A large deviation of the SM value of the phase $\chi $ is only possible in
models that violates $3\times 3$ unitarity. More precisely, in models with
an up singlet quark, $\chi $ can be order $\lambda $. Moderates effects
appear in $B_{d}^{0}\rightarrow \phi K_{S}$, more spectacular effects can be
present in CP asymmetries in the $B_{s}^{0}$ system. Correlated effects of
this scenario would be: rare top decays $t\rightarrow cZ$ at a rate
observable at LHC, production of a heavy top $T$ at LHC and important
deviations from 1 of $X_{tt}\sim \left\vert V_{tb}\right\vert ^{2}$,
measurable at ILC.

\section{Acknowledgments}

The present work has been done in different collaborations with J.A.
Aguilar-Saavedra, G.C. Branco, M. Nebot, M.N. Rebelo, and J.P. Silva. My
acknowledgment to all of them and special greetings to Gustavo Branco in his
60th birthday. This work has been partially founded by MEC of Spain under
FPA2002-00612.

\end{document}